\documentclass[aps, prl, 10 pt, notitlepage, nofootinbib, twocolumn, letterpaper, preprintnumbers, longbibliography, floatfix]{revtex4-1}

\pdfoutput=1
\usepackage{tikz}
\usepackage[compat=1.0.0]{tikz-feynman}

\usepackage[utf8]{inputenc}
\usepackage[normalem]{ulem}
\usepackage{graphicx}% Include figure files
\usepackage{dcolumn}% Align table columns on decimal point
\usepackage[colorlinks=true,allcolors=purple]{hyperref}
\usepackage{url}
\usepackage{enumerate}
\usepackage{slashed,multirow,relsize,soul,feynmp-auto,tikz}
\usepackage{color}
\usepackage{mathrsfs} % pretty maths
\usepackage{amsmath}
\usepackage{cancel}
\usepackage{bbold}
 \usepackage{mathrsfs}
\usepackage{braket}
\usepackage{physics}
\usepackage{multirow}
\usepackage[capitalize]{cleveref}
\usepackage{xspace}

\hypersetup{colorlinks,citecolor= nicegreen,linkcolor= nicered}
\definecolor{nicered}{rgb}{0.7,0.1,0.1}
\definecolor{nicegreen}{rgb}{0.1,0.5,0.1}
\begin{document}
\def\Carleton{Department of Physics, Carleton University, Ottawa, Ontario K1S 5B6, Canada}

\title{Neutrino Self-interaction and Weak Mixing Angle Measurements}
\author{Yue Zhang}
\affiliation{\Carleton}
%\email{yzhang@physics.carleton.ca}
%\date{\today}
\begin{abstract}
Neutrino self-interaction with a larger ``Fermi constant'' is often resorted to for understanding various puzzles of our universe. We point out that a light, neutrinophilic scalar particle $\phi$ through radiative correction leads to an energy-scale dependence in the neutrino-$Z$-boson gauge coupling. The driver behind this phenomenon is a large separation between the mass scales of $\phi$ and additional heavy particles needed for gauge invariance. This is a generic effect insensitive to details of the UV completion. We show that the running can change the $Z\nu\bar\nu$ coupling by several percent and affect the measurement of weak mixing angle through neutrino neutral-current processes. We discuss the interplay between the running of the $Z\nu\bar\nu$ coupling and $\sin^2\theta_W$ in various experimental observables. It is possible to disentangle the two effects with more than one precise measurement.
\end{abstract}
\maketitle

The self-interaction of active neutrinos is a blind spot among the tests of the Standard Model. Its strength is allowed to be much bigger than that dictated by the weak interaction if the self-interaction occurs through the exchange of a light, neutrinophilic mediator. At the cosmic frontier, the presence of self-interacting neutrinos appears favorable by recent analysis for improving the fit to data compared to the standard $\Lambda$CDM model~\cite{Kreisch:2019yzn, He:2023oke, Camarena:2024zck, Pal:2024yom}.
Neutrino self-interaction is also well motivated for enabling the production of sterile neutrino dark matter via early universe oscillation mechanism~\cite{DeGouvea:2019wpf, Kelly:2020pcy, Kelly:2020aks, Chichiri:2021wvw, Benso:2021hhh, An:2023mkf} which is otherwise excluded by astrophysical probes~\cite{DES:2020fxi}.
There have been numerous investigations of neutrino self-interaction at various frontiers, including rare decay processes~\cite{Barger:1981vd, Berryman:2018ogk, Blinov:2019gcj, Brdar:2020nbj, Lyu:2020lps, Dev:2024twk}, accelerator neutrino experiments~\cite{Kelly:2019wow, Kelly:2021mcd, Coyle:2022bwa}, high-energy colliders~\cite{deGouvea:2019qaz}, ultra-high-energy neutrino telescopes~\cite{Ioka:2014kca, Ng:2014pca, Ibe:2014pja, Kamada:2015era, Esteban:2021tub}, and core-collapse supernova~\cite{Shalgar:2019rqe, Chang:2022aas, Fiorillo:2023ytr}.

In this Letter, we point out a novel role of neutrino self-interaction at the precision frontier of testing the electroweak theory. Through radiative corrections, the presence of a light, neutrinophilic scalar can generate a running effect of the $Z\nu\bar\nu$ gauge coupling at energy scales below the $Z$-pole.
We show that existing constraints allow this running effect to modify the low-energy neutrino scattering cross sections via neutral-current by as large as a few percent.
Currently, there are strong interests and experiments on the horizon that study neutrino-matter interaction as tests of the Standard Model~\cite{Balantekin:2022jrq, Klein:2022lrf, deGouvea:2022gut, Giunti:2022aea}. The neutral-current processes involve another percent-level scale-dependent quantity, the weak mixing angle. We discuss how the running of $Z\nu\bar\nu$ coupling can impact the extraction of the weak mixing angle value from experiments, and the interplay of different measurements that can be used to distinguish the two running effects.

\medskip
\noindent{\bf Radiative correction to electroweak couplings.}\\
We begin with a simplified model commonly used for neutrino self-interaction analysis,
\begin{equation}\label{eq:1}
    \mathcal{L} = \frac{\lambda}{2} \bar \nu^c \nu \phi + {\rm h.c.} \ ,
\end{equation}
where $\nu$ is the left-handed neutrino chiral field and $\nu^c$ is its antiparticle, the right-handed antineutrino, and both belong to $SU(2)$ doublets in the Standard Model. The scalar field $\phi$ is a gauge singlet and has no direct coupling to other Standard Model particles. It can be real or complex and our discussions will not depend on this choice. The neutrino bilinear field carries two units of lepton number and Eq.~\eqref{eq:1} is the only Lorentz-invariant way for them to couple to a scalar field without introducing right-handed neutrinos.

In this work, we are interested in the mass of $\phi$ below the weak scale and explore its radiative corrections to the neutrino neutral-current and charged-current couplings. 
The tree-level electroweak Lagrangian is 
\begin{equation}
\mathcal{L} = g^0_Z \bar \nu \gamma^\mu \mathbb{P}_L \nu Z_\mu 
+ g_W^0 \bar \ell \gamma^\mu \mathbb{P}_L \nu W^-_\mu + {\rm h.c.} \ ,
\end{equation}
where $\mathbb{P}_L=(1-\gamma_5)/2$, $g_Z^0 = g/(2\cos\theta_W)$ and $g_W^0=g/\sqrt{2}$. $g$ is the $SU(2)$ gauge coupling and $\theta_W$ is the weak mixing angle.

At one-loop level, the relevant diagrams are
\begin{align}\label{eq:3}
\begin{tikzpicture}
\begin{feynman}
\vertex [label=above:\(Z\)] (a) at (0,0);
\vertex (b) at (0.7,0);
\vertex (c) at (1.57,0.5);
\vertex (d) at (1.57,-0.5);
\vertex [label=right:\(\nu\)] (e) at (2.27,0.5);
\vertex [label=right:\(\bar\nu\)] (f) at (2.27,-0.5);
\feynmandiagram [inline=(a.base)] {
(a)  -- [photon, momentum'=\(q\)] (b),
(b) -- [fermion, edge label=\(\nu^c\)] (c),
(d) -- [scalar, edge label'=\(\phi\)] (c),
(d) -- [fermion, edge label=\(\nu^c\)] (b),
(c) -- [fermion] (e),
(d) -- [anti fermion] (f),
};
\end{feynman}
\end{tikzpicture} 
\hspace{0.5cm}
\raisebox{0.65cm}{
$\equiv i g_{Z}^0 \Gamma_{Z\nu\bar\nu}^{(1)}(q^2)$ \ ,
} \\
\feynmandiagram [inline=(a.base), horizontal=a to d] {
  i1 [particle=\(\nu\)] -- [fermion, momentum'=\(p\)] a -- [fermion, edge label'=\(\nu^c\)] b -- [fermion] i2 [particle=\(\nu\)],
  a -- [scalar, half left, looseness=1.5, edge label=\(\phi\)] b,
}; 
\equiv -i\Sigma_\nu(p) \ , \nonumber
\end{align} 
where $\Gamma_{Z\nu\bar\nu}(q^2)$ and $\Sigma_\nu(p)$ are the amputated vertex correction and neutrino self-energy diagrams. 
$q^\mu$ is the momentum running through the $Z$ field.
Because $\phi$ is neutrinophilic, there are no corrections to the $W\ell\nu$ vertex or the charged lepton self-energy. The above is all one can get out of model \eqref{eq:1}.
The loops are equal to
\begin{eqnarray}\label{eq:4}
&&\hspace{-0.3cm}\Gamma_{Z\nu\bar\nu}(q^2, m_\phi^2) = -\frac{|\lambda|^2}{32\pi^2} 
\left\{ \frac{2}{\varepsilon'} - 1 + 2\int_0^1 dx \int_0^{1-x}dy \right. \nonumber\\
&&\hspace{-0.3cm}\left.\times\text{\small $\left[\ln \left(\frac{\mu^2}{(1-x-y)m_\phi^2-xy q^2}\right) + \frac{x y q^2}{(1-x-y)m_\phi^2-xy q^2} \right]$} \right\} \ ,\nonumber\\
&&\hspace{-0.3cm}\Sigma_\nu(m_\phi^2) = -\frac{|\lambda|^2}{32\pi^2} \left( \frac{2}{\varepsilon'} + \log \frac{\mu^2}{m_\phi^2} + \frac{1}{2}\right) \cancel{p} \ ,
\end{eqnarray}
where $2/\varepsilon' = 2/(4-d)-\gamma + \ln(4\pi)$ is used to regularize ultraviolet (UV) divergence in dimensional regularization and $\mu$ is the renormalization scale. In this calculation, we put the external neutrinos on mass shell. 

The corresponding radiative corrections to the electroweak couplings can be calculated as
\begin{equation}\label{eq:5}
\begin{split}
    \delta g_{Z}(q^2, m_\phi^2) &= g_{Z}^0 \left( \Gamma_{Z\nu\bar\nu}(q^2,m_\phi^2)  + \frac{\partial{\Sigma_\nu(m_\phi^2)}}{\partial{\cancel{p}}} \right) \ , \\
    \delta g_{W}(m_\phi^2) &= g_{W}^0 \frac{1}{2} \frac{\partial{\Sigma_\nu(m_\phi^2)}}{\partial{\cancel{p}}} \ .
\end{split}    
\end{equation}

The UV divergences in $\Gamma_{Z\nu\bar\nu}$ and $\partial{\Sigma_\nu}/\partial{\cancel{p}}$ add up rather than cancel each other in $\delta g_Z$. Unlike regular Yukawa interactions, $\phi$ in the Lagrangian \eqref{eq:1} turns a neutrino into an antineutrino, resulting in an extra minus sign for the $Z$-$\nu^c$ coupling in the triangle diagram in \eqref{eq:3}. Similarly, $\delta g_{W}$ only receives contribution from $\partial{\Sigma_\nu}/\partial{\cancel{p}}$ and is clearly UV divergent.

\medskip
\noindent{\bf Implications of gauge invariance, UV finiteness, and decoupling theorem.}\\
The UV divergences found above indicate that the model in Eq.~\eqref{eq:1} cannot be extrapolated to arbitrarily high scales. By assuming $\phi$ to be a singlet, it fails to respect the Standard Model gauge invariance. 

It is not feasible to assign $\phi$ a charge under $SU(2)\times U(1)$ which would open up the $Z\to \phi \phi^*$ decay channel for the light $\phi$ considered here. 
The latter makes an order-one correction to the invisible $Z$ decay width and is ruled out by the LEP-II experiment~\cite{ALEPH:1993pqw, OPAL:1994kgw, L3:1998uub}.

A phenomenologically viable UV completion of Eq.~\eqref{eq:1} was discussed in~\cite{Berryman:2018ogk}. In addition to the gauge singlet scalar $\phi$, a $SU(2)$ triplet scalar $T$ with hypercharge +1 is introduced. The interacting Lagrangian reads 
\begin{equation}\label{eq:triplet}
    \mathcal{L}_{\rm UV} = y_1 \bar L^c T L + y_2 H^T T^\dagger H \phi + {\rm h.c} \ .
\end{equation}
In the first term, the electric neutral component of the scalar triplet $T^0$ couples to $\bar \nu^c \nu$. After electroweak symmetry breaking, the second term generates a mixing between $\phi$ and $T^0$. Integrating out the triplet yields Eq.~\eqref{eq:1} and the relation $\lambda \simeq y_1 y_2 v^2/M_T^2$ where $v$ is the Higgs condensate. 
In this concrete model, the triplet $T$ field couples to neutrinos and charged leptons at the most fundamental level. It is straightforward to verify that the UV divergences from various vertex and self-energy diagrams indeed cancel completely.
The mixing between $\phi$ and $T^0$ is an infrared effect and cannot alter physics in the UV or the cancellation.

The notion about UV finiteness in the above example is not a coincidence but the consequence of the Ward identity. It holds on very general grounds. For any gauge-invariant, renormalizable extension of the Standard Model (BSM) that can accommodate Eq.~\eqref{eq:1}, the new particles would only contribute to the gauge couplings' beta functions through the vacuum polarization diagrams and the contribution freezes below the heavy particle mass threshold. The rest of radiative corrections to the gauge couplings must sum up to be finite.

This observation can allow one to write down the $Z\nu\bar\nu$ and $W\ell\nu$ gauge couplings in a model-independent way. Consider a generic UV completion of Eq.~\eqref{eq:1} where $\phi$ is the only new particle below the weak scale. All other BSM particles are heavy and $M$ is used to denote their mass scale. The above UV finiteness argument ensures that after including both $\phi$ and heavy loop contributions to $g_Z, g_W$, all $1/\varepsilon'$ terms must cancel away. Along with it, the renormalization scale $\mu$ will be replaced by $M$, which is now a physical quantity. 

For low momentum transfers $|q^2| \ll M^2$,
the one-loop renormalized gauge couplings take the form
\begin{widetext}
\begin{equation}\label{eq:6}
g_Z(q^2, m_\phi^2, M^2)
= g_Z^0 - \frac{g_Z^0|\lambda|^2}{32\pi^2} \left\{ \ln\frac{M^2}{m_\phi^2} + 2\int dx dy \left[\ln \left(\frac{M^2}{(1-x-y)m_\phi^2-xy q^2}\right) + \frac{x y q^2}{(1-x-y)m_\phi^2-xy q^2} \right] + c_Z\right\} \ ,
\end{equation}
\end{widetext}
and
\begin{equation}\label{eq:7}
g_W(m_\phi^2, M^2) = g_W^0 - \frac{g_W^0|\lambda|^2}{64\pi^2} \left(\ln\frac{M^2}{m_\phi^2} + c_W \right) \ .
\end{equation}

Most crucially for the following discussion, new physics at mass scale $M$ only serves the role for UV finiteness but has no impact on the momentum transfer $q^2$ dependence of the renormalized gauge couplings. Indeed, the $q^2$ dependence of Eq.~\eqref{eq:6} is identical to that from the light $\phi$ contribution in Eq.~\eqref{eq:4}.
This can be understood because heavy new physics cannot propagate at low energies. Their $q^2$-dependent contributions manifest as contact operators suppressed by powers of $q^2/M^2$, as dictated by the decoupling theorem~\cite{Appelquist:1974tg}.

The constants $c_Z$ and $c_W$ in Eqs.~(\ref{eq:6}, \ref{eq:7}) are order-one numbers and depend on details of the UV completion.
In the $SU(2)$ triplet scalar extension discussed above, the decoupling theorem sets a stronger requirement that not only the UV divergent terms but also finite terms should cancel between the $\phi$ and $T$ loop contributions if both are heavy. This can be used to fix $c_Z=-3/2$, $c_W=0$. More details are given in the appendix.

\medskip
\noindent{\bf Running $Z\nu\bar\nu$ coupling below weak scale.}\\
In the Standard Model, the $Z\nu\bar\nu$ gauge coupling does not run below the electroweak scale. In contrast, the renormalized $g_Z$ calculated in Eq.~\eqref{eq:6} has a logarithmic dependence on the momentum transfer $q^2$ and runs between $M$ and $m_\phi$ scales.
If the scalar $\phi$ mediating neutrino self-interaction is the only BSM particle lighter than the $Z$ boson, it generates a novel energy scale dependence in $g_Z$ at low energies. This running effect is enabled thanks to the separation of $m_\phi$ and $M$ mass scales.

The energy-scale dependence of $g_Z$ can be probed with neutrino neutral-current processes that exchanges the $Z$-boson at low energies.
The rest of this paper will focus on neutrino-matter scattering experiments where $q^2<0$.
We define $Q = \sqrt{-q^2}$ to be the experimental energy scale.
It is useful to examine the asymptotic behaviors of renormalized $g_Z$ as a function of $Q$,
\begin{equation}
\label{eq:asumptotic}
    \frac{g_Z}{g_Z^0} = 1 - \frac{|\lambda|^2}{32\pi^2}\times \left\{ \begin{array}{ll}
    \ln \frac{M^2}{Q^2} + \ln \frac{M^2}{m_\phi^2} +2+c_Z \ , &\hspace{0.0cm} Q \gg m_\phi \\
    2\ln \frac{M^2}{m_\phi^2} + \frac{3}{2} + c_Z \ , &\hspace{0.0cm} Q \ll m_\phi
    \end{array} \right.
\end{equation}

For $Q$ varying across large energy scales, we can resum the leading-log contributions and derive the renormalization group (RG) equation for $g_{Z}$,
\begin{equation}\label{eq:9}
    \frac{d g_{Z}}{d \ln Q} = \frac{|\lambda|^2}{16\pi^2} g_{Z} \ .
\end{equation}
The anomalous dimension for the $g_Z$ running, $|\lambda|^2/(16\pi^2)$, is a positive number, implying that $g_{Z}$ decreases for lower energy processes (smaller $Q$).

To solve for $g_{Z}$ using Eq.~\eqref{eq:9}, we still need an experimental input at a particular scale. It is a natural choice to use the $Z$ pole. The presence of heavy new physics in the UV completion of Eq.~\eqref{eq:1}, if not far above the electroweak scale, could modify the global fit to electroweak precision observables. Here we assume such an effect is small and $g_{Z}(M_Z)$ takes the best-fit value in the Standard Model.
This leads to a useful formula that works for $Q \gg m_\phi$,
\begin{equation}\label{eq:10}
    g^2_{Z}(Q) \simeq g^2_{Z}(M_Z) \left( \frac{Q}{M_Z}\right)^{{|\lambda|^2}/{8\pi^2}} \ .
\end{equation}

\begin{figure}[t]
    \centering
    \includegraphics[width=1\linewidth]{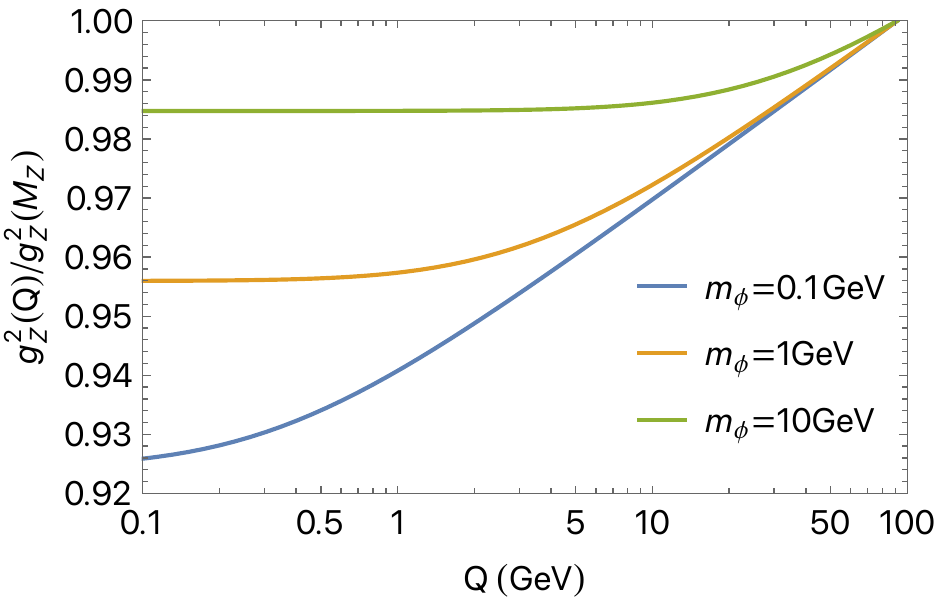}
    \caption{Running of the $Z\nu\bar\nu$ coupling with physical energy scale $Q$ in the presence of a light scalar $\phi$ that mediates neutrino self-interaction.}\label{fig:1}
\end{figure}

FIG.~\ref{fig:1} depicts the energy scale dependence of  $g_{Z}^2$ in the neutrino self-interaction model considered in this work, for three choices of $m_\phi$.
The coupling $\lambda$ is set to unity.
The curves are made using the one-loop result, Eqs.~\eqref{eq:6}, and they are insensitive to the value of $M$ as long as it is above $M_Z$.
For large $Q\gg m_\phi$, they numerically agree well with the RG-based formula Eq.~\eqref{eq:10}.
The running effect ceases after crossing the $m_\phi$ threshold.

We find that even for $m_\phi$ above GeV scale, which passes most existing constraints on the neutrinophilic scalar~\cite{Berryman:2022hds}, the change in $g_{Z}^2$ can be as large as a few percent level. The amount of change is comparable to the scale dependence of the weak mixing angle, another running effect to be discussed below.

It is also useful to remark that our discussions so far have ignored the neutrino flavors. In general, the running of $g_{Z}$ needs not be flavor universal because the neutrino self-interaction introduced in Eq.~\eqref{eq:1} can be a source of lepton flavor universality violation.

\medskip
\noindent{\bf Impact on $\sin^2\theta_W$ measurements with neutrinos.}\\
The weak mixing angle, or $\sin^2\theta_W$, is a fundamental parameter of the electroweak theory. At tree level, it is related to the $W/Z$-boson mass ratio. Including radiative corrections, $\sin^2\theta_W$ obtains an energy-scale dependence with the leading contribution from loop-generated $\gamma$-$Z$ kinetic mixing~\cite{Erler:2004in}. Running from the $Z$ pole down to the QCD scale, the Standard Model predicts $\sin^2\theta_W$ to grow by about 3\%~\cite{Kumar:2013yoa} which is consistent with most experimental probes. 

Existing low-energy limits that resort to electron-hadron and electron-electron interactions will not be affected by the $g_{Z}$ running effect found above. However, measurements of $\sin^2\theta_W$ using neutrino-matter scattering processes will do, because the neutral-current cross sections not only depends on $\sin^2\theta_W(Q)$ (in the $Z$-matter coupling) but also is proportional to $g_{Z}^2(Q)$. If one tries to interpret experimental data by simply assuming the $Z$-pole value of $g_{Z}^2(M_Z)$, its running effect would be attributed to that of $\sin^2\theta_W$, and the result would shift from the Standard Model prediction. The sign of the shift depends on the target of neutrino scattering. We will discuss several cases.

\begin{figure}[t]
    \centering
    \includegraphics[width=1\linewidth]{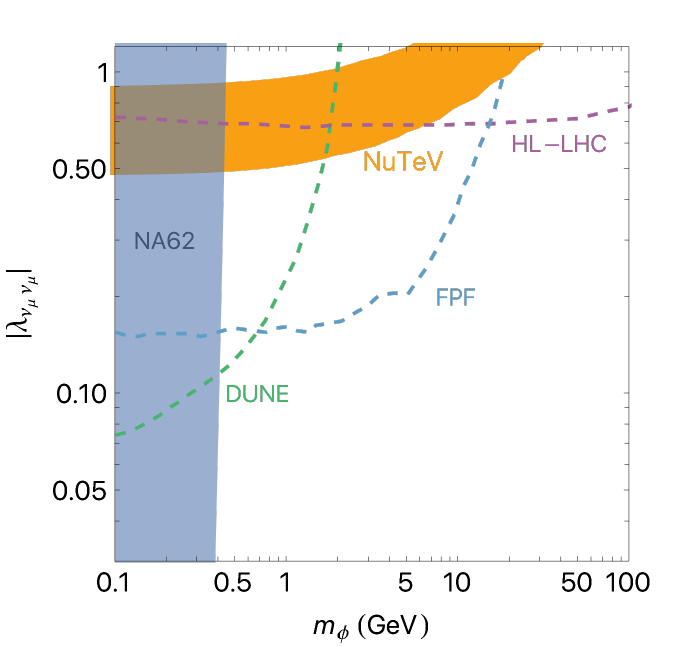}
    \caption{Parameter space of simplified model with a light, neutrinophilic scalar $\phi$. The orange shaded region can account for the NuTeV anomaly using the running effect of $g_Z$ while keeping the weak mixing angle consistent with the Standard Model predicted value. Blue shade region is excluded by the $K\to\mu\bar\nu + {\rm invisible}$ measurement by NA62~\cite{NA62:2021bji}. We also show several proposals to probe the model using the DUNE near detector~\cite{Kelly:2019wow}, the Forward Physics Facility~\cite{Kelly:2021mcd}, and ATLAS/CMS detectors at high-luminosity LHC~\cite{deGouvea:2019qaz}.}\label{fig:2}
\end{figure}

{\it Neutrino DIS. }
The first measurement of $\sin^2\theta_W$ using neutrino deep-inelastic scattering (DIS) was done by the NuTeV experiment~\cite{NuTeV:2001whx}. It shoots a $\nu_\mu$ or $\bar\nu_\mu$ beam onto the target at energy scale $Q\simeq 4.4$\,GeV. NuTeV measures the ratios of inclusive neutral-current to charged-current interaction cross sections and found they are lower than the Standard Model predictions (by few percent), leading to a higher $\sin^2\theta_W$~\cite{Loinaz:2002ep}. While it is commonly appreciated that the NuTeV anomaly could be attributed to hadronic uncertainties~\cite{Bentz:2009yy, Brodsky:2004qa}, for the discussions here it can serve an example to demonstrate the impact of scale dependence in $g_Z$ on the weak mixing angle measurement.
The $Q$-dependent quantities in the Paschos-Wolfenstein parameter~\cite{Paschos:1972kj} form the combination,
\begin{equation}\label{eq:12}
    g_{Z}^2(Q) \left[1-2\sin^2\theta_W(Q)\right] \ .
\end{equation}
Using Eq.~\eqref{eq:10}, we find that $g_{Z}^2(Q)$ runs in the correct direction to explain the deficit in the neutral-current events alone. The orange shaded region FIG.~\ref{fig:2} shows the parameter space that neutrino self-interaction allows the model in Eq.~\eqref{eq:1} to be consistent with NuTeV within $2\sigma$ level (we use data from Figure 1 of~\cite{Kumar:2013yoa}), while still keeping $\sin^2\theta_W(Q)$ as predicted in the Standard Model. We also show a recent constraint from exotic kaon decay measurement by the NA62 experiment~\cite{NA62:2021bji} which sets a lower limit on the viable $\phi$ mass~\cite{offshellphi}, as well as several proposals to further cover the parameter space at near-future experiments.
The novel $Q$ dependence of neutrino DIS (Eq.~\eqref{eq:12}) can be further examined with the LHC neutrinos at the Forward Physics Facility~\cite{MammenAbraham:2023psg}.

{\it CE$\nu$NS. }
The coherent elastic neutrino-nucleus scattering (CE$\nu$NS) is another channel to probe the running of $g_Z$. The CE$\nu$NS process has already been detected by COHERENT~\cite{COHERENT:2017ipa} and XENONnT~\cite{XENON:2024ijk, FeiGao2024}. More experiments are on the horizon to explore the CE$\nu$NS process with higher precision~\cite{Canas:2018rng, Scholberg:2020pjn}. The CE$\nu$NS cross section depends on $g_Z$ and $\sin^2\theta_W$ as~\cite{Goodman:1984dc, Formaggio:2012cpf}
\begin{equation}
    g_{Z}^2(Q) \left[N-Z(1-4\sin^2\theta_W(Q))\right]^2  \ ,
\end{equation}
where the square bracket is nuclear weak charge. Assuming isoscalar target with equal number of protons and neutrons ($N=Z$), the cross section is proportional to \begin{equation}
g_{Z}^2(Q)\sin^4\theta_W(Q) \ .
\end{equation}
The decrease of $g_{Z}^2$ at low energies could be interpreted as smaller weak mixing angle than the Standard Model value at the corresponding energy scale. Interestingly, this is an opposite effect to that for neutrino DIS.

{\it $\nu_\mu (\bar\nu_\mu)$--electron scattering. } Recent works make projections for measuring the weak mixing angle at accelerator neutrino experiments like DUNE and SBND with electron target~\cite{deGouvea:2019wav, Alves:2024twb}. These experiments can be tuned to run with either neutrino or antineutrino beam. The $\nu_\mu (\bar\nu_\mu)$-electron scattering cross sections have different parametrical dependences. In the $E_\nu\gg m_e$ limit,
\begin{equation}
    \begin{split}
        \sigma_{\nu_\mu e\to \nu_\mu e} &\propto g_{Z}^2(Q) \left[3 - 12\sin^2\theta_W(Q) + 16\sin^4\theta_W(Q) \right] \ , \\
        \sigma_{\bar\nu_\mu e\to \bar\nu_\mu e} &\propto g_{Z}^2(Q) \left[1 - 4\sin^2\theta_W(Q) + 16\sin^4\theta_W(Q) \right]\ .
    \end{split}
\end{equation}
For $\sin^2\theta_W$ around 0.23, the first square bracket is a decreasing function of $\sin^2\theta_W$, whereas the second one is an increasing function.
If neutrino self-interaction is at work and reduces $g_{Z}^2(Q)$ at low energies, it will lead to a suppression in both $\sigma_{\nu_\mu e\to \nu_\mu e}$ and $\sigma_{\bar\nu_\mu e\to \bar\nu_\mu e}$. If one tried to attribute this to the running of the weak mixing angle, they would reach opposite conclusions by using neutrino and antineutrino beams -- the former channel calls for a higher $\sin^2\theta_W(Q)$ but the latter lower. 

All the neutrino scattering cross sections discussed above are proportional to $g_Z^2$ but they depend on $\sin^2\theta_W$ in different ways.
With more than one precise measurement, one will be able to tell their running effects apart.

\medskip
\noindent{\bf Conclusion.}\\
Neutrino self-interaction via a light mediator particle is of great interest to address puzzles from the sky. 
We work on a simple extension of the Standard Model with a neutrinophilic scalar $\phi$ and calculate its radiative correction to the electroweak couplings. We derive a novel energy-scale dependence in the value of $Z\nu\bar\nu$ coupling below the $Z$-boson mass. This effect is granted by a large separation of $m_\phi$ and mass scale of other heavy particles in gauge invariant UV completions.
Thanks to the elusive nature of neutrinos, their couplings to the neutrinophilic mediator are still allowed to be sizable and the corresponding running effect of the $Z\nu\bar\nu$ coupling is allowed to be as large as several percent.
It affects the neutrino scattering cross sections via neutral-current interactions, and in turn measurements of the weak mixing angle, another scale-dependent quantity. We discuss the interplay of the two running effects and point out that they can be differentiated by combining measurements.
There is rich opportunity to disclose novel neutrino self-interaction at upcoming neutrino experiments.

\medskip
\noindent{\bf Acknowledgment.}\\
I thank Kevin Kelly and Miha Nemev\v{s}ek for useful discussions and comments on a draft.
This work is supported by a Subatomic Physics Discovery Grant (individual) from the Natural Sciences and Engineering Research Council of Canada.

\bibliographystyle{apsrev4-1}
\bibliography{ref}{}

\appendix
\section{Appendix}
In this appendix we derive the constants $c_Z, c_W$ in the renormalized gauge couplings (Eqs.~(\ref{eq:6}, \ref{eq:7})), for the model with a $SU(2)$ triplet scalar (Eq.~\eqref{eq:triplet}) as the UV completion of Eq.~\eqref{eq:1}. The discussion is based on the decoupling theorem.

First, we set the parameter $y_2=0$. In this case, the mixing between $\phi$ and $T^0$ vanishes, and $\phi$ completely decouples from other particles. All the BSM radiative corrections are from the interaction between $T$ and the lepton doublet $L$. If we give the whole $SU(2)$ triplet a gauge-invariant and large mass $M$, all its effects on the Standard Model as the low-energy effective theory must be suppressed by inverse powers of $M$.

Next, we turn on a nonzero $y_2$ and in turn the $\phi$-$T^0$ mixing. 
This will reshuffle radiative corrections (and the UV divergences, which will still cancel) to various loop diagrams involving $\phi$ and $T$.
The mixing will enter through various vertex Feynman rules.
To proceed along the above argument, we are after the case where the $\phi$-$T^0$ mixing does not vanish in the decoupling limit. It is possible to realize this situation by examining the $\phi$-$T^0$ mixing mass matrix,
\begin{equation}\label{eq:app}
    \begin{pmatrix}
        \mu_T^2 & y_2 v^2 \\
        y_2 v^2 & \mu_\phi^2
    \end{pmatrix} \ ,
\end{equation}
where $\mu_T^2$ and $\mu_\phi^2$ are from the quadratic $T$ and $\phi$ potential terms before the mixing. The useful decoupling limit for our discussion here will be sending both $\mu_T$ and $\mu_\phi$ to infinity while keeping them equal to each other. In this case, $\phi$ and $T^0$ are nearly degenerate and the mass matrix is always diagonalized with a $(\pi/4)$ mixing angle. As a result, the induced $\phi$-neutrino coupling does not vanish, neither do the diagrams in Eq.~\eqref{eq:3}. 

By sending both $\phi$ and $T$ mass to infinity, the decoupling of their joint effects should continue to work. It means if one includes all the radiative correction contributions from $\phi$ and $T$, no only their UV divergence but also the finite terms must cancel away, up to $q^2/M^2$ terms. This argument provides a simple way to evaluate the (many) $T$-loop diagrams without needing to directly compute them, because we have already calculated the $\phi$ ones for general $\phi$ mass.
The heavy $T$ contributions to $g_Z$ and $g_W$ are given by
\begin{equation}
    \begin{split}
        \delta g_Z^{(T)} &= - \lim_{M^2\gg|q^2|} \delta g_{Z} (q^2, M^2) \\
        &= \frac{|\lambda|^2}{16\pi^2} \left( \frac{2}{\varepsilon'} - \ln\frac{\mu^2}{M^2} + \frac{1}{2} \right)  \ , \\
        \delta g_W^{(T)} &= - \lim_{M^2\gg|q^2|} \delta g_{W} (M^2) \\
        &= \frac{|\lambda|^2}{64\pi^2} \left( \frac{2}{\varepsilon'} - \ln\frac{\mu^2}{M^2} + \frac{1}{2} \right)  \ ,
    \end{split} 
\end{equation}
where the functions $\delta g_{Z, W}$ are given by Eq.~\eqref{eq:5} and
$M$ denotes the physical mass of $T$.

By dialing the new physics parameters, it is possible to make $\phi$ light while keeping $T$ still heavy. The $\phi, T^0$ couplings will change correspondingly, but the number of Feynman diagrams relevant for radiative corrections and the structure of the loop functions ($q^2$ and mass dependence) still hold as before.
As long as the mass of $T$ remains well above the momentum transfer, the above forms of $\delta g_{Z,W}^{(T)}$ will hold intact.
Adding them to the $\phi$ contributions yields Eqs.~(\ref{eq:6}, \ref{eq:7}) with $c_Z=-3/2, c_W=0$.

A consistency check is the $Q\ll m_\phi$ limit of Eq.~\eqref{eq:asumptotic}. With $c_Z=-3/2$, we find the radiative correction to $g_Z$ is proportional to $\ln(M^2/m_\phi^2)$. With $c_W=0$, the radiative correction of $g_W$ in Eq.~\eqref{eq:7} is proportional to the same logarithmic factor.
In the $(\pi/4)$ mixing scenario discussed above (with $\mu_T=\mu_\phi$), Eq.~\eqref{eq:app} gives $M^2=\mu_T^2 + y_2 v^2$ and $m_\phi^2 = \mu_T^2-y_2 v^2$. The logarithmic factor vanishes as $\sim v^2/\mu_T^2$ for very large $\mu_T$, fulfilling the desired decoupling behavior.

\end{document}